# An Efficient Admission Control Algorithm for Load Balancing In Hierarchical Mobile IPv6 Networks

Prof P.Harini

HOD of IT Department
StAnns College of Engineering & Technology
Chirala, AndhraPradesh, India
hariniphd@yahoo.com

Dr. O.B.V.Ramanaiah

Professor in CSE Department
JNTUCEA
Ananthpur,
AndhraPradesh, India

*Abstract*— **In hierarchical Mobile IPv6 networks, Mobility Anchor Point (MAP) may become a single point of bottleneck as it handles more and more mobile nodes (MNs). A number of schemes have been proposed to achieve load balancing among different MAPs. However, signaling reduction is still imperfect because these schemes also avoid the effect of the number of CN's. Also only the balancing of MN is performed, but not the balancing of the actual traffic load, since CN of each MN may be different. This paper proposes an efficient admission control algorithm along with a replacement mechanism for HMIPv6 networks. The admission control algorithm is based on the number of serving CNs and achieves actual load balancing among MAPs. Moreover, a replacement mechanism is introduced to decrease the new MN blocking probability and the handoff MN dropping probability. By simulation results, we show that, the handoff delay and packet loss are reduced in our scheme, when compared with the standard HMIPv6 based handoff.**

## I. INTRODUCTION

### A. Mobile IP

In the early days of the Internet's development a decision was made that Internet protocol (IP) addresses would represent both the topological location and identity of an end-host (RFC 791) [Pos81b]. While this decision simplified the Internet's conceptual addressing model and met the needs of early network deployments it has created difficulties for the development and deployment of truly mobile, IP-based applications and services. In the 1990s a network-layer 'Mobile IP' solution was first developed in the context of IP version 4 (IPv4) (RFC 3344) [Per02a]. During this same period of time, the Internet Engineering Task Force (IETF) began work on a new version of IP that has become known as IP version 6 (IPv6) (RFC 2460) [DH98]. Despite the fact that IPv6 addressing still maintains much of IPv4's semantic link on location and identity, experience with Mobile IPv4 allowed the IETF to integrate better support for Mobile IP1 into IPv6 [1].

The conventional IP addresses represent the host's identity and encode the host's topological location on the IP network [Pos81b] simultaneously. This problem is solved by the mobile IP address. By following the physical movement of the host's attachments, a host gives the result of entering into a secondary network with respect to the networks IP topology. Once if this takes place, a new IP address must be assigned to the host therefore the packets may be correctly routed to the host's new location. The transport layer connections are broken due to this type of moves which were active for the duration of the movement. This is because the hosts IP address is also used as a transport level and end point identity. So the packets are lost which are sent to the earlier IP address. In addition to this, when the hosts sends the IP packets at this moment then their prior peers will not know the new address.

When two IP addresses are established for mobile hosts then the mobile IP assists in this region of this problem. They are static 'home address and a transitory' care of addresses. The static 'home address' of the host is identified worldwide. The transitory 'care of address' identified temporarily when it is attached to the different parts of the network. However the mobile host is attached to the internet by its foreign address through dynamically managed IP-in-IP tunnels and particularly encoded packet forwarding rules and it is operated from its home address. As a result of this, the network layer mobility is supported by Mobile IP to be transparent towards all upper layers because just the address is being used by the transport layer. The non-mobile internet will function even if the mobile host environment applications are designed according to the conventional assumptions.

### B. Mobile IPv6

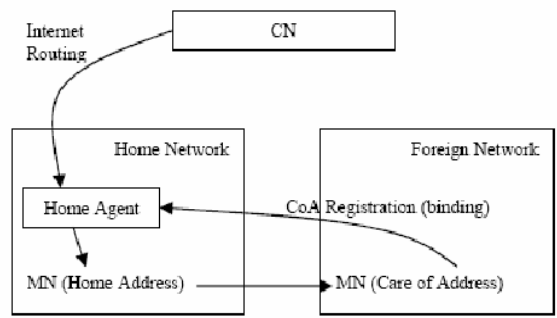

Figure 1. Mobile IPv6

Mobile IPv6 is designed to manage the Mobile Node (MN) movements in the IPv6 networks. Every MN is available to all other hosts with Mobile IPv6 through its home identity. In the MN's home network, a Home Agent (HA) is router. By maintaining the MN's current location information, HA stores the mobility of the MN. The last router in the foreign network is the Access Router (AR) which can forward the packets to MN. The idea of Foreign Agent (FA) does not exist in the mobile IPv6. AR takes part as a substitute for the FA except assigning its IP address (the MN's Care-of Address (CoA) would be the IP address).

A new CoA is obtained by the MN when it moves to a new sub network. The CoA is either configured using IPv6 stateless address auto configuration or stateful address configuration. Then the MN registers this CoA to the HA.

There are two possible ways for the MN and Correspondent Node (CN) to communicate when the MN is away from home: direct communication (Route Optimization) and tunneling via the HA. In Mobile IPv6, route optimization is fundamental. MN sends binding updates to CNs for direct communication. On the other hand, if the CN is a non-Mobile IPv6 node, the CN can communicate with the MN via tunneling through the HA. Packets from a CN destined to the MN's Home Address are intercepted by the HA and then tunneled to the MN's CoA, and vice versa (packets from the MN CoA are encapsulated and tunneled to the HA and forwarded to the CN).

*C. Hierarchical Mobile IPv6*

The network overhead and the increased signaling messages are introduced by the basic mobile IP when the MN varies its point of attachment to different networks frequently. A localized mobility management protocol for MN is given by the HMIPv6 which is an extension of basic Mobile IPv6. In order to handle mobility management a new conceptual entity called Mobility Anchor Point (MAP) is introduced by this scheme. MAP is a router or a set of routers. An exclusively reliable administration of a particular domain is maintained by this MAP. With the help of its common routable IP address (Figure 1), MAP connects the domain and serves the Internet. A regional care-of address (RCoA) and an on link care-of address (LCoA) are the two types of the MN addresses present in HMIPv6. The RCoA specifies a particular domain of the Internet and it is called as global address.

The MN transforms its LCoA when it is moved between the local networks inside a MAP domain. It also requires the new LCoA to register in a MAP on the local link. The MN changes both the address when it is moved from one MAP domain to the new MAP domain. Therefore, this needs registering new local LCoA and new RCoA to the new MAP. At this instant, the new MAP registers global RCoA to the MN's HA.

The MN has the capability of choosing between the basic mode and the extended mode when it is moved inside a MAP domain. In basic mode, for receiving the packets which are sent to MN inside the same domain, the MAP acts as the same as the local HA. The packets which are intended to the RCoA are received by the MAP and send them to the corresponding LCoA of the MN. The RCoA is the MAP's address in the extended mode. With the MN home address, the MAP keeps a

binding table with the current LCoA. When the MAP receives the packets which are intended to a MN then it

*D. Handoff in HIMPv6*

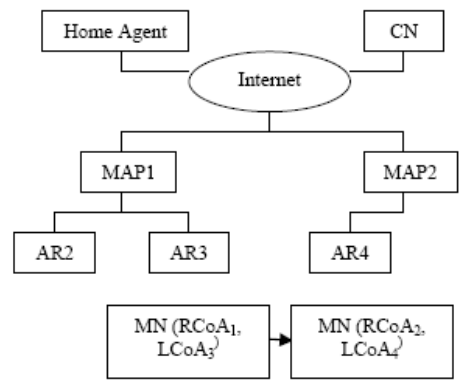

Figure 2. Inter-domain Handoff in HIMPv6

The following circumstances are the basis for the system for Inter-domain handoff: a regional Care-of Address RCoA1 and an on-link Care-of Address LCoA3 (Figure. 2) are present in the MN. When the CN sends packets to the MN, the packets will be sent through MAP1 to the MN's LCoA3 [3].

*E. Problems in HMIPv6*

Generally the performance of standard MIPv6 is always better than HMIPv6. Decreased signaling outside an MAP domain cannot compensate the increased signaling inside the MAP domain, when the number of CNs and the number of handoffs of each MN is low. Thus, an MN should decide whether it is effective to use HMIPv6 under certain conditions.

Instead, an MAP may become a single point of bottleneck as it handles more and more MNs, A number of schemes have been proposed to achieve load balancing among different MAPs. Among them, schemes based on multiple MAP levels are proposed. The multiple MAP levels share the traffic load and MAP selection algorithm is based on the MN's speed. However, signaling reduction is still imperfect because these schemes also avoid the effect of the number of CN's.

In some schemes, threshold- based admission control algorithms are proposed to avoid overload at particular MAPs. Each MAP set a threshold to limit the number of serving MNs. Such algorithms can only manage the balancing of MN, but not the balancing of the actual traffic load, since CN of each MN may be different.

This paper proposes an efficient load balancing scheme along with a replacement mechanism for HMIPv6 networks. It introduces an algorithm for enabling an MN to determine whether it is suitable to use HMIPv6. The admission control algorithm is based on the number of serving CNs achieves actual load balancing among MAPs. Moreover, the replacement mechanism is introduced to decrease the new MN blocking probability and the handoff MN dropping probability.

## II. RELATED WORK

Nakajima.N [4], have proposed a Robust Hierarchical Mobile IPv6 (RH-MIPv6), which provides incorrect tolerance and strength in mobile networks. In RH-MIPv6 [5], a mobile node (MN) registers primary (P-RCoA) and secondary (S-RCoA) regional care of addresses to two different MAPs (Primary and Secondary) at the same time. They have developed a method to permit the mobile node or correspondent node (CN) for the detection of the failure of primary MAP and modify their attachment from primary to secondary MAP.

Moon et al. [6] proposed a method to lessen the handover delays. This is done by cooperating with the information of layer 2 and in addition by performing some functions of the mobile node on the access router. They have recommended a slight more competent and optimized mobility scheme. This method has combined appropriately the advantages of earlier proposed schemes.

Pekka Nikander et al [7] discussed the design rationale behind the MIPv6 Route Optimization Security Design.IPv6 (MIPv6) allows a Mobile Node to talk directly to its peers while retaining the ability to move around and change the currently used IP address. This mode of operation is called Route Optimization (RO), as it allows the packets to traverse a shorter route than the default one through the Home Agent. This is potentially dangerous, since a malicious host might be able to establish false bindings, thereby preventing some packets from reaching their intended destination, diverting some traffic to the attacker, or flooding third parties with unwanted traffic.

Aisha Hassan Abdalla Hashim et al. [8] described how mobile node can change its point of attachment from one access router to another in Mobile IPv6 (MIPv6). Hierarchical Mobile IPv6 (HMIPv6) is one of them that are designed to reduce the amount of signaling required and to improve handover speed for mobile connections. This is achieved by introducing a new network entity called Mobility Anchor Point (MAP). This presents a comparative study of the Hierarchical Mobility IPv6 and Mobile IPv6 protocols and we have narrowed down the scope to micro-mobility (intra-domain).

Md. Mohiuddin Khan et al, [9] discussed about Mobile IP and the proposed handoff schemes, their pros and cons and some areas where they can be improved. The reasons for handoff latency are examined and the schemes are discussed by their genres. They observe that combining different methods results in a better performance. They buffer issues like cross-layer design and context awareness improve the handoff latency but also bring overhead to them. Later those packets are delivered. This study demonstrated the fundamental points of concern related with efficient handoff designs.

Pyung-Soo Kim, and Yong Jin Kim [10] the new fast vertical handover scheme is proposed for the hierarchical Mobile IPv6 (HMIPv6) to optimize and enhance the existing fast vertical handover HMIPv6 (FVH-HMIPv6) in heterogeneous wireless access networks. The recently standardized IEEE 802.21 Media Independent Handover Function (MIHF) is adopted for the proposed FVH-HMIPv6. Firstly, the Media Independent Information Service (MIIS) is extended by including new L3 information to provide domain prefixes of heterogeneous neighboring mobility anchor points (MAPs), which is critical to the handover performance of proposed FVH-FMIPv6 with MIHF. Secondly, the operation procedure for the proposed scheme is described in detail.

Li Jun ZHANG and Samuel PIERRE [11] presents a comprehensive performance analysis of Fast handover for Hierarchical Mobile IPv6 (F-HMIPv6) using the fluid-flow and random walk mobility models. Location update cost, packet delivery cost and total cost functions are formulated based on the proposed analytical models. They investigate the impact of several wireless system factors such as user velocity, user density, mobility domain size, session-to-mobility ratio on these costs, and present some numerical results.

In our previous work we [12] recommend a novel cross-protocol design approach for the DAD problem. It uses information from current routing protocol traffic. This utilizes Passive Duplicate Address Detection (PDAD) to identify duplicate addresses. Using PDAD, a node examines incoming routing protocol packets to develop hints about address controversies. Since it is independent of the routing protocol, it produces nearly no protocol overhead and reduces initial delay.

In our previous work we [13] propose an Enhanced Hierarchical Mobile IPv6 (E-HMIPv6) architecture based on a novel cross-layer/cross-protocol design approach. A new node, called Helper node, is adopted in our (E-HMIPv6) architecture which can assist the Mobile Station (MS) to obtain the (regional care of address) RCoA and LCoA (local care of address) and execute the DAD procedure for proving the unique. The MS can still transmit data to the Correspondent Node (CN) during the pre-handoff procedure. Thus it is evident that the latency of handoff in the projected protocol is lower than in the traditional HMIPv6 and MIPv6. For the DAD problem, we use the Passive Duplicate Address Detection (PDAD) mechanism to identify duplicate addresses. In order to develop the hints about the address controversies, a node examines the incoming routing protocol packets using PDAD. Being independent of the routing protocol the PDAD does not create any protocol overhead and reduces initial delay.

## III. ADMISSION CONTROL

Let us consider the scenario depicted in Figure.3 as an example. Three MAPs MAP1, MAP2 and MAP3 are placed. Each MAP consists of access routers. Let MAP2 contains two access routers AR1 and AR2. MN2 is moving from AR1 to AR2, resulting in handoffs and the process of admission control. Here, we assume that the MAPs that the two MNs registered previously are no longer valid.

AR2 achieves MAP options through route advertisements. When MN2 attaches to AR2's link, it is admitted as per the admission control algorithm.

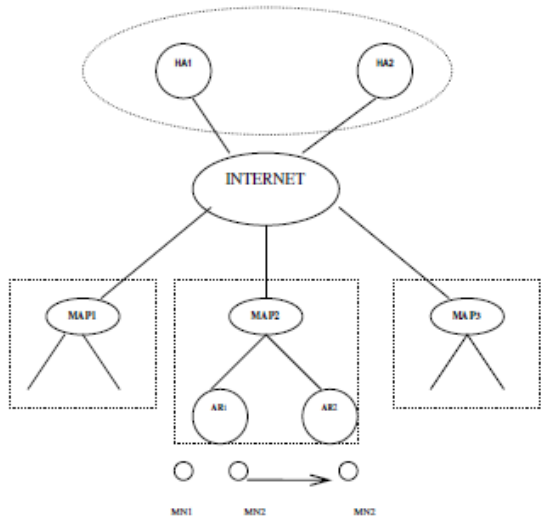

Figure 3: Architecture of the Proposed Scheme

## A. Classification of MN

In the proposed scheme, incoming MNs are classified into two types:

- new MN

- handoff MN

(1) New MN: The MN performing the initial binding updates BU) to the MAP (e.g. when an MN is turned on).

(2) Handoff MN: When an ongoing MN moves into a new MAP domain, the MN sends a local BU message towards the new MAP to complete local registration. Since the handoff MN is registered, it is probably communicating with one or more CNs. Thus, the handoff MN should have higher priority than the new MN.

An MAP should determine whether the received BU message comes from a handoff MN or a new MN using the following steps:

1. A flag A is added to the existing BU message, and it is set if the MN is in the ready state when it sends a BU message.

2. If the flag A in BU is set, the MN is regarded as a handoff MN. Otherwise, it is regarded as a new MN.

3. A ready state timer is controlled by MN to know if it is in ready or idle state. If an MN in idle state sends or receives data, the ready timer is initialized and the MN changes to the ready state. Every time the MN sends or receives data, the ready timer is reset. If the MN does not send or receive data until the ready timer expires, the MN returns to the idle state.

In the proposed admission control algorithm for the two types of MN's, two threshold values are maintained as

- $N_{thr}$ - threshold value for new MN

- $H_{thr}$ - threshold value for handoff MN

Let $tot_{cn}$ denotes the total number of CNs that a MAP currently serves.

Normally $H_{thr}$ is equal to the capacity of a MAP which is the maximum number of CNs that the MAP can serve.

The admission control algorithm is defined as:

**Admission Control Algorithm**

1. Determine $tot_{cn}$

2. *if* $tot_{cn} \leq N_{thr}$ then

    *Admit* both $MN_{new}$ *and* $MN_{hand}$

3. *if* $N_{thr} < tot_{cn} \leq H_{thr}$ then

    *Admit* $MN_{hand}$ and reject $MN_{new}$

4. *if* $H_{thr} < tot_{cn}$ then

    Reject both $MN_{new}$ and $MN_{hand}$

Where $MN_{admit}$ indicates the MNs which can be admitted under the above conditions, $MN_{hand}$ denotes the handoff MNs and $MN_{new}$ denotes the new MNs

## B. Replacement Mechanism

When a new MN or a handoff MN cannot be accepted, it should not be blocked or dropped directly. The MAP will choose an MN from existing MNs to be replaced. If the number of connecting CNs ($con_{cn}$) of an MN is equal to or larger than that of the incoming MN, it becomes a candidate to be replaced. The MAP sends binding acknowledgement (BA) to the chosen MN. The BA message contains an error code with the reason "Insufficient resources". Then, the replaced MN performs the following MAP selection algorithm to choose another suitable MAP among the remaining MAPs.

In proposed scheme, MN maintains a table containing information on available MAPs until new router advertisement arrives. Thus, it enables the MN to choose different MAPs.

**MAP Selection Algorithm**

1. *for* each $MAP_i$

    1.1 *Calculate* the ratio Y

        $Y = con_{cn} / tot_{cn}$

    1.2. *Calculate* the combined measure W

        $W = \alpha(Y + S)$

    *Where* S - speed of MN and $\alpha$ - constant

    1.3. *if* $W < T_{map}$ then

        *Choose* $MAP_i$

    *Where* $T_{map} - threshold$ value

2. *end* for

## IV. EXPERIMENTAL RESULTS

### A. Simulation Setup

We use NS2 to simulate our proposed (E-HMIPv6) architecture. In our simulation, the channel capacity of mobile hosts is set to the same value: 2 Mbps. We use the distributed coordination function (DCF) of IEEE 802.11 for wireless LANs as the MAC layer protocol. It has the functionality to notify the network layer about link breakage. The following table (Table.1) summarizes the simulation settings. The CBR traffic is established from CN to MS, and the bandwidth and latency for every link between every two components are also specified in this scenario.

TABLE I. SIMULATION SETTINGS

| No. of Nodes | 20 |
|---|---|
| Area Size | 1000 X 1000 |
| Mac | 802.11 |
| Simulation Time | 50 sec |
| Traffic Source | CBR |
| Packet Size | 512 |
| Speed | 5,10,15,20 |
| Transmission range | 75m |
| Routing Protocol | AODV |

For the simulation, we make use of Hierarchical Mobile IP (HMIP) implementation, which has implemented in Columbia IP Micro-mobility Software (CIMS). It supports micro mobility protocols for instance Hawaii, cellular IP, and HMIP extension meant for the ns-2 network simulator based on version 2.1b6. We have additionally included MAP functionality to provide regional registration with the existing CIMS implementations.

The simulation has carried out using the network topology shown in Figure 4. It consists of MAP1, MAP2 and MAP3. Each MAP contains two access routers.

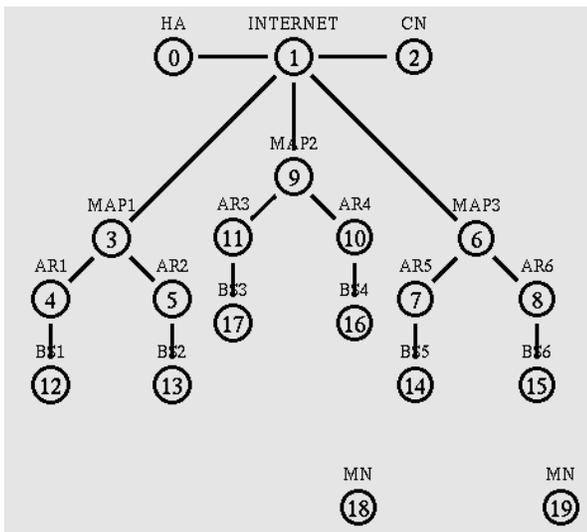

Figure 4. Network Topology

Initially the mobile node MN19 was in MAP3 in the domain AR6. During the simulation we perform intra and inter domain handoff on MN19.

Initially, at time t1, the mobile node performs intra domain handoff by moving from AR6 to AR5 within MAP3. Next at time t2, it start moving towards AR4 from AR5, thus by performing inter domain handoff. At time t3, it moves from AR4 to AR3, within MAP2. Finally at time t4, it moves back to AR1, once again performing inter domain handoff.

We evaluate the performance of our scheme based on the following parameters.

- **Handoff Latency:** The handoff latency is defined as the time interval from last packet received form serving BS to and new packet received from target BS.

- **Throughput:** The number of packets received at the MS.

### B. Results

#### A. Varying Rate

Initially we vary the rate of traffic as 0.1Mb, 0.2Mb….0.5Mb.

Figure 5 shows the packets received for AC-HMIPv6 and E-HMIPv6 schemes. From the figure, we can see that the throughput increases when the rate increases. Due to its load balancing scheme, AC-HMIPv6 has more throughput when compared with E-HMIPv6.

Figure 6 shows the handoff delay for E-HMIPv6 and AC-HMIPv6 based schemes. Clearly the handoff delay for (AC-HMIPv6) is significantly less when compared with E-HMIPv6.

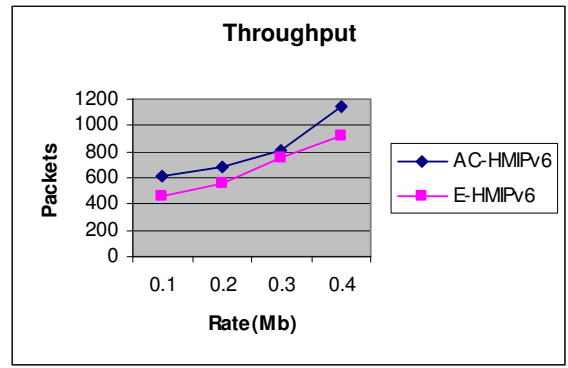

Figure 5. Rate Vs Throughput

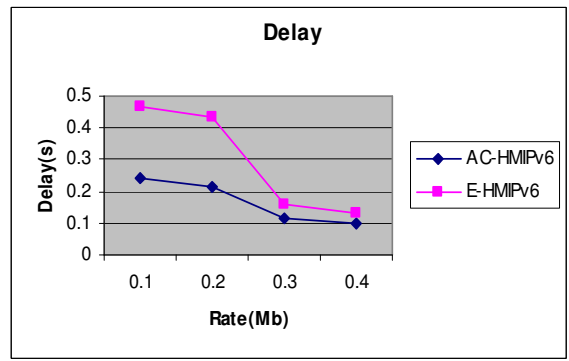

Figure 6. Rate Vs Handoff Delay

#### B. Varying Speed

Next we vary the speed of the MN as 5, 10, 15 and 20 m/s.

Figure 7 shows that the throughput decreases when the speed of the mobile increases since it has to perform the admission control and MAP selection algorithms within a short period. From Fig 7, we can see that the throughput is once again less in the case of E-HMIPv6 scheme when compared with our AC-HMIPv6.

Figure 7 shows that the delay decreases as the speed of the mobile increases since it has to perform the MAP selection algorithms within a short period. We can see that AC-HMIPv6 scheme has low delay when compared with E-HMIPv6 scheme.

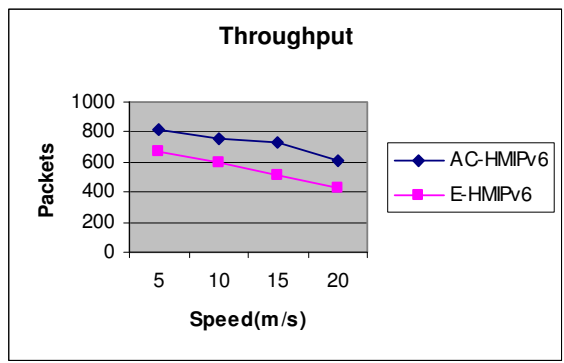

Figure 7. Speed Vs Throughput

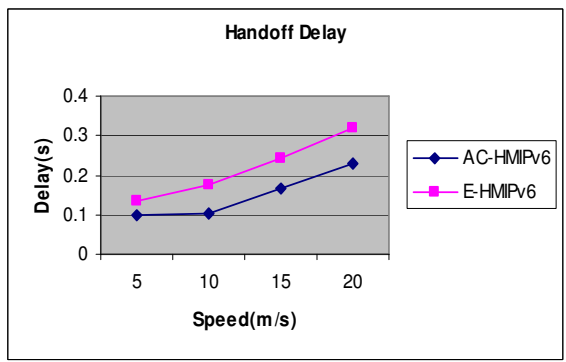

Figure 8. Speed Vs Handoff Delay

Thus, the AC-HMIPv6 based architecture with our implementation strategies adapts all scenarios perfectly.

## V. CONCLUSION

In this paper we have proposed an efficient admission control algorithm along with a replacement mechanism for HMIPv6 networks. In the proposed scheme, incoming MNs are classified as new MN and handoff MN. It consists of a technique by which a MAP can determine whether the received BU message comes from a handoff MN or a new MN. The admission control algorithm is based on the number of serving CNs and achieves actual load balancing among MAPs. Moreover, a replacement mechanism is introduced to decrease the new MN blocking probability and the handoff MN dropping probability. By simulation results, we have shown that, the handoff delay and packet loss are reduced in our scheme, when compared with the standard HMIPv6 based handoff. As a future work, we like to include recovery mechanisms for packet loss occurred due to handoff. Also we need to develop a rate control mechanism based on the bandwidth availability of the new access point.

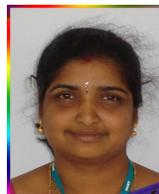

**Prof. P. Harini** M.Tech.(Remote Sensing),M.Tech. (CSE), [Ph. D. (Mobile Computing)]. I obtained my M.Tech. (Remote Sensing) in 1997 & M.Tech. (CSE) in 2003 from JNTU, Masab Tank, Hyderabad. I worked as a Research Associate in JNTU, Masab Tank, Hyderabad in Remote Sensing Department for 01 year, 05 years worked as a Assistant Professor in QIS College of Engineering, Ongole and 01 year worked as a Associate Professor in Rao & Naidu Engineering College, Ongole. At present I am working as Professor & Head of the Computer Science and Engineering Department in St. Ann's College of Engineering & Technology, Chirala.

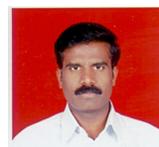

**Dr. O.B.V. Ramanaiah** received Ph.D. in Computer Science from University of Hyderabad in 2005, M. Tech in Computer Science and B. Tech in Computer Science & Engineering . My Total Teaching Experience is16 Years with Total Research Experience of 12 Years. I have 3 Publications in International Conference Proceedings and 2 International Journals .I have Visited USA for paper presentation in an IEEE Conference, ITCC 04, held during April 5-7, 2004. I am providing Research Guidance to 8 students and have organized many Refresher Courses.